# Evaluation of an information security management system at a Mexican higher education institution

Rúsbel Domínguez Domínguez, Omar Flores Laguna, Jair A. del Valle López

Facultad de Ciencias Empresariales y Jurídicas, Universidad de Montemorelos, Montemorelos, México.

rusbel@um.edu.mx, oflores@um.edu.mx, jdelvalle@um.edu.mx

## 1 Introduction

Information is a very important asset for institutions. Therefore, it is essential to protect information to avoid risks. The storage and processing of information requires a high level of responsibility to minimize cyber threats [1, 2].

The information security management system (ISMS) manages risk based on methodologies of international norms and standards (ISO/IEC 27000) that help to minimize threats and expose vulnerable targets of information systems in organizations. The implementation of these techniques allows managers to make the best decisions. That said, it is understood that if an institution fails in information security management, the integrity of its data will be compromised and in turn could have financial implications [3, 4].

Increasing cybersecurity threats have forced organizations to implement a set of security methodologies in their information management systems. The execution of these practices has served as a countermeasure to current information risks in organizations [3, 5].

The degree of information risk management is an important mechanism in the strategic management of corporations, which can sometimes be decisive in coordinating business actions, maintaining operations and providing business continuity [6, 7].

ISO/IEC 27000 is a series of standards that allow the implementation of security components to protect the information systems that form part of the most important assets of organizations. This standard describes the guidelines that organizations must have to guarantee the availability, integrity, and authenticity as well as the confidentiality of information [8].

The information security management system (ISMS) is a standardized and systematised process based on the ISO/IEC 27002 standards which makes it possible to analyse and assess the risks of an organizations IT assets and by means of which the availability, integrity and reliability of a company's information management is guaranteed [9].

ISMS is defined as a management system used to establish and maintain an environment to safeguard information. ISMS considers the ongoing maintenance of procedures and processes for operating information technology security, in which actions are contemplated to identify information security vulnerabilities, implement strategies that contribute to the reduction of risks, get to know the needs, take proper measure of the results, and improve the protection strategies through a cycle of continuous improvement and risk mitigation [10].

ISMS procedures, policies and instructions are specific to each information system which aims to protect the information assets in an institution [11].

The operations in the implementations of ISMS are processes that are generally characterized by having several actions executed in a determined order and requiring a succession of resources, equipment, facilities, personnel, as well as a series of accesses to obtain a result [12].

Several authors [8, 9, 13] mention that the implementation of these standard-based processes allows organizations to guarantee the availability, integrity, authenticity, and confidentiality of information through policies and regulations.

It is important to mention the benefits of proper risk management in educational institutions. Researchers executed an ISMS model in basic education centres in Colombia. They examined the critical assets in the areas of the academic secretariats of several educational institutions based on the NTC ISO/IEC 27001:2013 standard. The model managed to meet the requirements set by the standard, which helps to ensure the availability, integrity, and the confidentiality of student information [13].

On the other hand, researchers observed that in an Asian university the increase of threats and damages to confidentiality, integrity and availability of information increased exponentially. By implementing an information management project, they became aware of the importance of the security of their information assets, which allows the protection of students' personal data, research, and technological infrastructure by adopting best practices in information security management based on the ISO/IEC 27001:2013 standard [14].

Furthermore, a survey of various economic sectors showed that an educational institution leads in compliance followed by large companies and then small and medium-sized enterprises. However, it states that educational institutions, despite leading in compliance are far from achieving ISO 27001 certification [15].

The objectives of this research were to: (a) to assess possible differences in the knowledge, training, commitment, effective management, and total average ISMS as perceived by the administrators and staff of the Information Technology Department (ITD) of a university in north-eastern Mexico. This was perceived by administrators and ITD personnel of a university in north-eastern Mexico. (b)Assess possible differences in the knowledge, training, commitment, effective management, and the mean total ISMS average between the technical value of perceived risk level by administrators of a university in north-eastern Mexico. (c)Assess possible differences in the knowledge, training, commitment, effective management, and the mean total ISMS (made up of the four variables), as perceived by ITD personnel, and the technical value of the risk level of a university in north-eastern Mexico. (d)Assess possible differences in the knowledge, training, commitment, effective management of an ISMS and the technical scale of perceived risk level by ITD personnel of a university in north-eastern Mexico. (e)Assess possible differences in knowledge, training, commitment, effective management and mean total ISMS among administrators. The technical value was obtained in previous research that is in the process of publication [16] by analysing vulnerabilities in the wireless network of the same university under study in north-eastern Mexico.

## 2 Methodology

This study has a quantitative approach and it is cross-sectional as the information was collected in a single time and correlational - comparative, as it aims to compare the groups. The population consisted of 81 subjects (66 administrators and 15 ITD personnel). Those evaluated were employed in an administrative role at the university and staff of the Information Technology Department (ITD).

A non-random convenience sample (managers who were willing to participate) was drawn, the total sample obtained was 48 administrators and 15 ITD personnel from a university in north-eastern Mexico. To make the comparisons, three groups of managers were formed according to classifications of administrative staff made by other [17], the classification was as follows: (a) first-line manager, (b) middle management and (c) top management.

### 2.1 The instrument

To assess the level of implementation of an Information Security Management System (ISMS), a scale was developed considering some ISO/IEC 27002 criteria. And was validated in research that is in the process of publication [18]. This survey consists of 24 items divided into four factors: (a) organization policies and regulations (PR1 to PR9), (b) privacy (P10 to P14), (c) integrity (I15 to I20) and (d) authenticity (A21 to A24). The composite reliability using the Omega coefficient for the factors were as follows: (a) policies and regulations was equal to .947, (b) authenticity was equal to .932, (c) integrity was .936 and (d) privacy was equal to .892. Thus, it is shown that the reliability is acceptable on all factors.

The scale assesses the ISMS, however, for this research the instrument was adapted to conduct a self-perception of the participants on the following four aspects: (a) knowledge, (b) training, (c) commitment and (d) effective management.

The technical value was obtained in previous research that is in the process of being published [16] through the analysis of vulnerabilities in the wireless network of the same university under study in north-eastern Mexico which is made up of 300 Access Points (APs) of the Aruba Networks brand. The technical value obtained was 5.9, categorised at a medium risk level for the security of the wireless network according to the National Vulnerability Database (NVD).

## 4 Analysis of results

*4.1 Descriptive and inferential statistics*

For data analysis, IBM SPSS Statistics version 25 software was used. Within the descriptive statistics, the mean, standard deviation, skewness, kurtosis, and the significance of Shapiro Wilk was calculated for each variable of each of the hypotheses raised. The threshold for skewness and kurtosis was used as the range -2 to 2 for the univariate normality of the items [19] and the Shapiro Wilk test for significance.

For the analysis of the null hypotheses, the student's *t-test* for independent samples was used for hypothesis Ho1, the one-sample *t-test* for hypotheses Ho2 and Ho3, as well as the one-factor Anova for Ho4. The rejection criterion for the null hypotheses was for *p*-significance values less than or equal to .05.

For the effect size of the *t-tests* Cohen's *d* was used, the reference values were: (a) $d=.20$ small effect size, (b) $d=.50$ medium effect size and (c) $d=.80$ large effect [20].

For the effect size of the one-factor Anova *F-tests* the reference values were: $f=.10$ small effect size, $f=.25$ medium effect size and $f=.40$ [20].

4.2 Null hypothesis testing one

Null hypothesis (Ho1): no significant difference in perception of knowledge, training, commitment, and effective administration and the mean total ISMS (made up of the four variables), as assessed by IT administrators and IT staff at a university in north-eastern Mexico.

*4.2.1 Comparison of first-line manager and ITD personnel*

Table 1 shows that the variable's knowledge, training, commitment, effective management, and ISMS are in the range of -2 to 2 in skewness and kurtosis for both first-line manager and ITD staff. In some of the variables, when applying the Shapiro Wilk test, *p*-values of less than .05 are observed, but very close to this value, so it was decided to take the variables as normal distributions.

*Table 1. Descriptive statistics for first-line manager and ITD staff*

| Variables | Type of respondent | Mean | Standard deviation | Skewness | Kurtosis | Significance Shapiro-Wilk |
|---|---|---|---|---|---|---|
| Knowledge | First-line manager | 4.670 | 2.642 | 0.310 | -1.571 | .054 |
| | ITD personnel | 6.949 | 1.446 | -0.732 | 0.120 | .538 |
| Training | First-line manager | 4.089 | 2.520 | 0.571 | -1.177 | .062 |
| | ITD personnel | 6.286 | 1.613 | -0.895 | 0.236 | .130 |

| | | | | | |
|---|---|---|---|---|---|
| Commitment | First-line manager | 5.194 | 2.811 | -0.018 | -1.876 | .039 |
| | ITD personnel | 7.185 | 1.602 | -1.347 | 1.204 | .024 |
| Administration | First-line manager | 4.979 | 2.921 | -0.038 | -1.923 | .025 |
| | ITD personnel | 6.824 | 1.510 | -1.366 | 1.266 | .013 |
| ISMS | First-line manager | 4.733 | 2.631 | 0.129 | -1.647 | .053 |
| | ITD personnel | 6.811 | 1.470 | -1.286 | 0.972 | .025 |

Table 2 shows the comparison of the first-line manager and ITD staff groups where a significant difference was observed in knowledge ($p=.009$), training ($p=.011$), commitment ($p=.032$), effective management ($p=.049$) and ISMS ($p=.018$). Large effect sizes were observed for knowledge, training, commitment, and ISMS, while a medium effect size was observed for effective management.

*Table 2. Comparison of first-line manager and ITD personnel*

| Variable | Statistic $t$ | Significance | Decision | $d$ |
|---|---|---|---|---|
| Knowledge | -2.832 | .009 | Reject Ho | 1.070 (Large) |
| Training | -2.747 | .011 | Reject Ho | 1.038 (Large) |
| Commitment | -2.303 | .032 | Reject Ho | 0.870 (Large) |
| Administration | -2.100 | .049 | Reject Ho | 0.794 (Medium) |
| ISMS | -2.580 | .018 | Reject Ho | 0.975 (Large) |

*4.2.2 Comparison of middle management and DTI personnel*

Table 3 shows that the skewness and kurtosis are at the required threshold; also, most of the variables when applying the Shapiro Wilk test, the $p$-values were greater than .05, so it was decided to take the variables as normal distributions.

*Table 3. Descriptive statistics for middle management and DTI personnel*

| Variable | Type of respondent | Mean | Standard deviation | Skew-ness | Kurtosis | Significance Shapiro-Wilk |
|---|---|---|---|---|---|---|
| Knowledge | Middle mgmt. | 4.899 | 2.783 | 0.510 | -1.392 | .063 |
| | ITD personnel | 6.949 | 1.446 | -0.732 | 0.120 | .538 |
| Training | Middle mgmt. | 4.384 | 3.063 | 0.489 | -1.358 | .082 |
| | ITD personnel | 6.286 | 1.613 | -0.895 | 0.236 | .130 |
| Commitment | Middle mgmt. | 5.507 | 2.926 | 0.347 | -1.399 | .160 |
| | ITD personnel | 7.185 | 1.602 | -1.347 | 1.204 | .024 |
| Administration | Middle mgmt. | 5.321 | 2.828 | 0.384 | -1.196 | .240 |
| | ITD personnel | 6.824 | 1.510 | -1.366 | 1.266 | .013 |
| ISMS | Middle mgmt. | 5.028 | 2.856 | 0.479 | -1.396 | .090 |

|  |  |  |  |  |  |
|---|---|---|---|---|---|
| ITD personnel | 6.811 | 1.470 | -1.286 | 0.972 | .025 |

Table 4 shows the inferences in the comparison of the middle management and ITD staff groups, which shows a significant difference in the knowledge variable ($p$=.024); and a large effect size.

*Table 4. Comparison of middle management and ITD personnel*

| Variable | Statistic $t$ | Significance | Decision | $d$ |
|---|---|---|---|---|
| Knowledge | -2.446 | .024 | Reject Ho | 0.925 (Large) |
| Training | -2.056 | .053 | Retain Ho | 0.777 (Medium) |
| Commitment | -1.881 | .074 | Retain Ho | 0.711 (Medium) |
| Administration | -1.754 | .095 | Retain Ho | 0.663 (Medium) |
| ISMS | -2.077 | .051 | Retain Ho | 0.785 (Medium) |

*4.2.3 Comparison of top management and ITD personnel*

Table 5 shows that the skewness and kurtosis are at the required threshold; also, most of the variables when applying the Shapiro Wilk test, the $p$-values were greater than .05, so it was decided to take the variables as normal distributions.

*Table 5. Descriptive statistics for top management and ITD personnel*

| Variable | Type of respondent | Mean | Standard deviation | Skewness | Kurtosis | Significance Shapiro-Wilk |
|---|---|---|---|---|---|---|
| Knowledge | Top mgmt. | 6.520 | 2.310 | -0.723 | -0.674 | .167 |
|  | ITD personnel | 6.949 | 1.446 | -0.732 | 0.120 | .538 |
| Training | Top mgmt. | 6.201 | 2.323 | -0.545 | -0.525 | .525 |
|  | ITD personnel | 6.286 | 1.613 | -0.895 | 0.236 | .130 |
| Commitment | Top mgmt. | 7.029 | 2.345 | -0.816 | -0.392 | .087 |
|  | ITD personnel | 7.185 | 1.602 | -1.347 | 1.204 | .024 |
| Administration | Top mgmt. | 6.469 | 2.246 | -0.557 | -0.424 | .266 |
|  | ITD personnel | 6.824 | 1.510 | -1.366 | 1.266 | .013 |
| ISMS | Top mgmt. | 6.554 | 2.224 | -0.730 | -0.170 | .420 |
|  | ITD personnel | 6.811 | 1.470 | -1.286 | 0.972 | .025 |

Table 6 shows the inferences in the comparison of the top management and ITD staff groups, in which no significant difference was observed in any of the variable´s.

*Table 6. Comparison of senior management and ITD personnel*

| Variable | Statistic *t* | Significance | Decision | *d* |
|---|---|---|---|---|
| Knowledge | -.590 | .561 | Retain Ho | 0.223 (Small) |
| Training | -.112 | .911 | Retain Ho | 0.042 (Very small) |
| Commitment | -.205 | .839 | Retain Ho | 0.077 (Very small) |
| Administration | -.491 | .628 | Retain Ho | 0.185 (Very small) |
| ISMS | -.360 | .722 | Retain Ho | 0.136 (Very small) |

4.3 Null hypothesis tests two

Null hypothesis (Ho2): no significant difference between the level of knowledge, training, commitment, and effective administration and the mean total ISMS (made up of the four variables), as perceived by administrators, and the technical value of risk level at a university in north-eastern Mexico.

In the analysis of this hypothesis, the one-sample *t-test* was performed. The criterion for rejection of the null hypotheses was for *p*-significance values less than or equal to .05.

For the effect size of the *t-tests* Cohen's *d* was used, the reference values were: (a) $d=.20$ small effect size, (b) $d=.50$ medium effect size and (c) $d=.80$ large effect size [20].

The technical value of the vulnerability analysis on the wireless network of the same university under study was 5.9, categorised at a medium risk level according to the National Vulnerability Database (NVD) [16].

Normality tests were performed on hypothesis Ho1 (see Table 1, Table 3 and Table 5).

*4.3.1 Comparison of first-line manager and technical value*

Table 7 shows the descriptive statistics in the comparison between the first-line manager and the technical value of 5.9, in which negative mean differences are observed in the variables, meaning that the mean in the variables is lower than the technical value.

*Table 7. Descriptive statistics of first-line manager and value of technician*

| Variable | Type of respondent | Mean | Technical value | Difference of means | Standard deviation |
|---|---|---|---|---|---|
| Knowledge | First-line manager | 4.670 | 5.9 | -1.230 | 2.642 |
| Training | First-line manager | 4.089 | 5.9 | -1.811 | 2.520 |
| Commitment | First-line manager | 5.194 | 5.9 | -.707 | 2.811 |
| Administration | First-line manager | 4.979 | 5.9 | -.921 | 2.921 |
| ISMS | First-line manager | 4.733 | 5.9 | -1.167 | 2.631 |

Table 8 shows the inferences in the comparison of the first-line manager and technical value groups, in which there is a significant difference in the training variable ($p=.019$), with a medium effect size.

*Table 8. Comparison of first-line manager and technical value*

| Variable | Statistic *t* | Significance | Decision | *d* |
|---|---|---|---|---|
| Knowledge | -1.742 | .105 | Retain Ho | 0.465 (Small) |
| Training | -2.689 | .019 | Reject Ho | 0.719 (Medium) |
| Commitment | -.941 | .364 | Retain Ho | 0.251 (Small) |
| Administration | -1.180 | .259 | Retain Ho | 0.315 (Small) |
| ISMS | -1.660 | .121 | Retain Ho | 0.443 (Small) |

*4.3.2 Comparison of middle management and technical value*

Table 9 shows the descriptive statistics in the comparison between the middle management group and the technical value of 5.9, in which negative mean differences are observed in the variables, meaning that the mean in the variables is lower than the technical value.

*Table 9. Descriptive statistics of middle management and technician value*

| Variable | Type of respondent | Mean | Technical value | Difference of means | Standard deviation |
|---|---|---|---|---|---|
| Knowledge | Middle mgmt. | 4.899 | 5.9 | -1.001 | 2.783 |
| Training | Middle mgmt. | 4.384 | 5.9 | -1.517 | 3.063 |
| Commitment | Middle mgmt. | 5.507 | 5.9 | -.393 | 2.926 |
| Administration | Middle mgmt. | 5.321 | 5.9 | -.579 | 2.828 |
| ISMS | Middle mgmt. | 5.028 | 5.9 | -.872 | 2.856 |

Table 10 shows the inferences in the comparison of the middle management groups and the technical value, in which no significant difference is observed in any of the variables.

*Table 10. Comparison of middle management and technical value*

| Variable | Statistic *t* | Significance | Decision | *d* |
|---|---|---|---|---|
| Knowledge | -1.346 | .201 | Retain Ho | 0.360 (Small) |
| Training | -1.853 | .087 | Retain Ho | 0.500 (Medium) |
| Commitment | -.502 | .624 | Retain Ho | 0.134 (Very small) |
| Administration | -.765 | .458 | Retain Ho | 0.204 (Small) |
| ISMS | -1.143 | .274 | Retain Ho | 0.305 (Small) |

*4.3.3 Comparative of top management and technical value*

Table 11 shows the descriptive statistics of the variables where positive mean differences were found in relation to the technical value, meaning that the means of the top management in each variable are above the technical value.

*Table 11. Descriptive statistics of top management and value of technician*

| Variable | Type of respondent | Mean | Technical value | Difference of means | Standard deviation |
|---|---|---|---|---|---|
| Knowledge | Top mgmt. | 6.520 | 5.9 | .619 | 2.310 |
| Training | Top mgmt. | 6.201 | 5.9 | .301 | 2.323 |
| Commitment | Top mgmt. | 7.029 | 5.9 | 1.129 | 2.345 |
| Administration | Top mgmt. | 6.469 | 5.9 | .569 | 2.246 |
| ISMS | Top mgmt. | 6.554 | 5.9 | .655 | 2.224 |

Table 12 shows the comparison of the top management and technical value groups, in which no significant difference is observed in any of the variables.

*Table 12. Comparison of top management and technical value*

| Variable | Statistic $t$ | Significance | Decision | $d$ |
|---|---|---|---|---|
| Knowledge | 1.003 | .334 | Retain Ho | 0.269 (Small) |
| Training | .484 | .636 | Retain Ho | 0.129 (Very small) |
| Commitment | 1.801 | .095 | Retain Ho | 0.481 (Small) |
| Administration | .949 | .360 | Retain Ho | 0.254 (Small) |
| ISMS | 1.101 | .291 | Retain Ho | 0.294 (Small) |

4.4 Null hypothesis test three

Null hypothesis (Ho3): no significant difference in the perception of knowledge, training, commitment, and effective administration and the mean total ISMS (made up of the four variables), as perceived by ITD personnel, and the technical value of the risk level of a university in north-eastern Mexico.

In the analysis of this hypothesis, a one-sample *t-test* was performed. The rejection criterion for the null hypotheses was for *p*-significance values less than or equal to .05.

Normality tests were performed on hypothesis Ho1 (see Table 1, Table 3 and Table 5).

*4.4.1 Comparison of the ITD personnel and technical value*

Table 13 shows the descriptive statistics of the variables where positive mean differences were found in relation to the technical value, meaning that the means of the DTI staff in each variable are above the technical value.

*Table 13. ITD personnel and technical value*

| Variable | Type of respondent | Mean | Technical value | Difference of means | Standard deviation |
|---|---|---|---|---|---|
| Knowledge | ITD personnel | 6.949 | 5.9 | 1.049 | 1.446 |
| Training | ITD personnel | 6.286 | 5.9 | .386 | 1.613 |
| Commitment | ITD personnel | 7.185 | 5.9 | 1.285 | 1.602 |
| Administration | ITD personnel | 6.824 | 5.9 | .924 | 1.510 |
| ISMS | ITD personnel | 6.811 | 5.9 | .911 | 1.470 |

Table 14 shows the comparison of ITD staff and technical value, in which significant differences were found in the variable's knowledge ($p=.018$), commitment ($p=.010$), effective management ($p=.039$) and ISMS ($p=.037$). In the variable training no significant differences were found ($p=.387$), a large effect size was observed for commitment and a medium effect size for knowledge, effective management, and ISMS.

*Table 14. Comparison of ITD personnel and technical value*

| Variable | Statistic $t$ | Significance | Decision | $d$ |
|---|---|---|---|---|
| Knowledge | 2.716 | .018 | Reject Ho | 0.726 (Medium) |
| Training | .895 | .387 | Reject Ho | 0.240 (Small) |
| Commitment | 3.000 | .010 | Reject Ho | 0.802 (Large) |
| Administration | 2.290 | .039 | Reject Ho | 0.612 (Medium) |
| ISMS | 2.319 | .037 | Reject Ho | 0.620 (Medium) |

4.5 Null hypothesis test four

Null hypothesis (Ho4): no significant difference in the perception of knowledge, training, commitment, and effective administration and the mean total ISMS (made up of the four variables), evaluated by administrators (first-line manager, middle management, and top management) of a university in north-eastern Mexico.

*4.5.1 Comparison between administrators*

When comparing managers, it was observed that the variables knowledge, training, commitment, effective management, and ISMS have a skewness and kurtosis between -2 and 2 in the three groups (first-line manager, middle management, and top management). In most of the variables, when applying the Shapiro Wilk test, $p$-values greater than .05 were observed, so it was decided to take the variables as normal distributions between the groups, therefore, the one-factor Anova *F-test* was applied.

When statistical inferences were drawn, the following results were found: knowledge ($F_{(2,39)} = 2.128$, $p= .133$), training ($F_{(2,39)} = 2.600$, $p= .087$), commitment ($F_{(2,39)} = 1.843$, $p= .172$), administration ($F_{(2,39)} = 1.186$, $p= .316$) y ISMS ($F_{(2,39)} = 2.005$, $p= .148$). As the $p$-values are greater than .05, the null hypothesis is retained for all variables, meaning that there is no significant difference in the groups on all variables.

## 5　Discussion

The purpose of this research was to know the degree of administrative knowledge, the degree of training of human resources, the degree of commitment of administrators and the degree of effective administration for information security risk based on ISO/IEC 27001.

Based on this study, it can be determined that in the Ho1 analysed for first-line manager and ITD staff there is a significant difference in relation to the variable's knowledge, training, commitment, effective management, and ISMS. In case of middle management and ITD staff, there is only significant difference in the knowledge variable. Top management and ITD staff do not show any significant difference.

About this result, it can be corroborated that administrative staff with a lower rank have more problems in making the best decisions in relation to the implementation of an ISMS, it should be noted that the first-line manager is the one who has more contact with the students and is the one who is less involved in the implementation of an ISMS.

Similarly, it was found that for Ho2 when comparing first-line man-manager with the test value there is no significant difference in most of the variables except for the training variable. For middle management and top management there is no significant difference in all variables. This suggests that the first-line managers, being the executors of the institution's planning, are not fully trained in the institution's information security efforts. This in turn prevents the generation of proposals for initiatives to implement an ISMS. With this shortcoming, it is possible that security breaches could be generated.

When assessing Ho3 by comparing the DTI staff with the technical value, a significant difference is observed in the variable's knowledge, commitment, effective management, and ISMS. What can be determined as a slightly elevated perception of the level of implementation of an ISMS in relation to the level of risk given by the technical value. DTI staff consider that they are well trained, have sufficient commitment, adequate knowledge, and correct effective administration, as the technical value of security is below the average for each variable.

It could be determined that Ho4 there was no significant difference between the groups of managers, in terms of knowledge, training, commitment, effective management and ISMS. However, a higher perception of top management was observed in the total means compared to the first-line manager and middle management groups.

Some researchers mention that implementing methodologies based on the ISO 27001 standards allows to establish a level of judgement in the security processes of an organization and to determine which process is not in place that may be necessary [21].

Similarly, proposing a solid model with security indicators can help the cybersecurity auditor to provide recommendations to increase the level of security and thus avoid security breaches [22, 23].

**Acknowledgments** The authors wish to thank César Augusto Puesán Frometa, Alejandro García Mendoza and Damaris Tarango Alvidrez, Writing - review & editing.